\documentclass[journal,twoside,web]{ieeecolor}
\usepackage{tmi}
\usepackage{cite}
\usepackage{amsmath,amssymb,amsfonts}
\usepackage{algorithmic}
\usepackage{graphicx}
\usepackage{textcomp}

\usepackage{listings}
\usepackage{xcolor}

\lstdefinestyle{pycode}{
	language=Python,
	basicstyle=\ttfamily\small,
	columns=fullflexible,
	keepspaces=true,
	showstringspaces=false,
	breaklines=true,
	frame=single,
	tabsize=4
}

\usepackage{multirow}
\usepackage{array}
\usepackage{hhline}

\usepackage{caption}
\usepackage{comment}

\def\BibTeX{{\rm B\kern-.05em{\sc i\kern-.025em b}\kern-.08em
		T\kern-.1667em\lower.7ex\hbox{E}\kern-.125emX}}
\begin{document}
	\title{Brain MR Image Synthesis with 3D Multi-Contrast Self-Attention GAN}
	\author{Zaid A. Abod, Furqan Aziz
		\thanks{The authors are affiliated with the School of Computing and Mathematical Sciences, University of Leicester, Leicester, United Kingdom. Email: zawaa2@leicester.ac.uk (Zaid A. Abod); fa311@leicester.ac.uk (Furqan Aziz).}
	}
	
	\maketitle
	
	\begin{abstract}
		Complete and high-quality multi-modal Magnetic Resonance Imaging (MRI) is essential for accurate neuro-oncological assessment, as each contrast provides complementary anatomical and pathological information. However, acquiring all modalities (e.g., T1c, T1n, T2w, T2f) for every patient is often impractical due to prolonged scan times, cost, and patient discomfort, potentially limiting comprehensive tumour evaluation. We propose 3D-MC-SAGAN (3D Multi-Contrast Self-Attention Generative Adversarial Network), a unified 3D multi-contrast synthesis framework that generates high-fidelity missing modalities from a single T2w input while explicitly preserving tumour characteristics. The model employs a multi-scale 3D encoder--decoder generator with residual connections and a novel Memory-Bounded Hybrid Attention (MBHA) block to capture long-range dependencies efficiently, and is trained with a WGAN-GP critic and an auxiliary domain classification head to produce T2f, T1n, and T1c volumes within a unified network. 
		To ensure anatomical and pathological fidelity, we incorporate a frozen 3D U-Net-based segmentation network that enforces a tumour-consistency constraint during training. A composite objective combining adversarial, reconstruction, perceptual, structural similarity, contrast-classification, and segmentation-guided losses further promotes both global realism and tumour-preserving structure.
		Extensive experiments on 3D brain MRI datasets demonstrate that 3D-MC-SAGAN achieves state-of-the-art quantitative performance and produces visually coherent, anatomically plausible contrasts with improved distributional realism. Importantly, the proposed method maintains tumour segmentation accuracy comparable to that achieved using fully acquired multi-modal inputs, highlighting its potential to reduce acquisition burden while preserving clinically meaningful information.
	\end{abstract}
	
	\begin{IEEEkeywords}
		GAN, Multimodality MRI Synthesis, Perceptual Loss, Residual Networks, Self-Attention, Tumour Segmentation
	\end{IEEEkeywords}

	\section{Introduction}
	\label{sec:introduction}
	\IEEEPARstart{M}{agnetic} Resonance Imaging (MRI) is integral to modern medical diagnostics due to its non-invasive acquisition of high-resolution soft-tissue images and its widespread use for studying neuroanatomy and diagnosing brain disorders \cite{ref1,ref2,ref3,ref4}. Since its development in the 1980s, it has significantly enhanced medical diagnostics \cite{ref5}. MRI operates by generating a strong magnetic field that aligns protons within biological tissues. This imaging modality is clinically valuable for evaluating neuropathologies affecting the brain’s visual pathway, including inflammation, demyelination, trauma, malignancy, and ischemia \cite{ref6}. 
	
	To comprehensively characterise brain structure and pathology, MRI is typically acquired using multiple imaging sequences, each highlighting different tissue properties. Common modalities include T1-weighted (T1n), T2-weighted (T2w), Fluid-Attenuated Inversion Recovery (T2f), and contrast-enhanced T1 (T1c), which provide complementary contrasts for detecting neurological disorders, particularly brain tumours \cite{ref8,ref9,ref10}. T1n provides high-resolution anatomical detail, with white matter appearing brighter than grey matter, facilitating the identification of tumours and vascular or inflammatory changes \cite{ref11}. T2w emphasises high-water-content processes such as edema and inflammation. T2f suppresses cerebrospinal fluid (CSF), enhancing the visibility of lesions near CSF spaces, and improves detection of periventricular and juxtacortical lesions \cite{ref11,ref7,ref10,ref12}. T1c is central to neuro-oncologic imaging, enabling improved detection of brain metastases, differentiation between enhancing tumour and non-enhancing edema, and accurate lesion delineation for monitoring and diagnosis \cite{ref3,ref6,ref13}. Integrating these complementary sequences improves tumour assessment and diagnostic accuracy \cite{ref14,ref15,ref16,ref17}. However, acquiring all modalities is often impractical due to long scan times, patient discomfort, motion artifacts, and resource constraints; in some cases, implanted medical devices may also preclude MRI examinations \cite{ref7,ref8,ref18}.
	
	Driven by recent advances in generative AI, research has increasingly focused on synthesising missing modalities using generative models rather than developing task-specific models that operate with incomplete inputs. Medical image synthesis involves learning a mapping from an observed source modality to an unobserved target modality. However, this task is inherently ill-posed, as the model must infer target images in the absence of target-modality observations \cite{ref20}. To address these practical constraints and resulting modality gaps, recent research has pursued computational approaches that infer unavailable image contrasts directly from the acquired data. Principal applications include cross-modality synthesis to translate between imaging sequences or modalities; modality completion for missing-image imputation; image enhancement and restoration such as super-resolution, denoising, and artifact reduction; and virtual dataset creation for augmentation and rare-case simulation, thereby supporting downstream analysis and clinical decision-making \cite{ref19,ref21,ref22,ref23}.
	
	Among the widely adopted computational approaches, generative adversarial networks (GAN) combined with autoencoder-based architectures have shown strong performance\cite{ref20,ref24,ref25,ref26,ref27,ref28,ref29}. Early models, such as pix2pix \cite{ref30} and CycleGAN \cite{ref31}, demonstrated the feasibility of cross-modality image translation \cite{ref2,ref8}. However, these approaches are limited to one-to-one mappings, requiring separate models for each modality pair. 
	Furthermore, their reliance on pixel-wise reconstruction losses, which tend to encourage intensity averaging, and PatchGAN discriminators that primarily enforce local realism can lead to global structural inconsistencies and poor preservation of subtle pathological features, such as tumour boundaries and tissue subregions critical for clinical analysis \cite{ref2,ref32}. Subsequent methods, including StarGAN-based variants \cite{ref33}, introduced multi-domain translation by conditioning a single generator-discriminator framework on domain labels with cycle-consistency constraints. This unified design enables efficient multi-modal synthesis without training multiple pairwise models. Nevertheless, maintaining pathological fidelity remains a major challenge: synthesised images must accurately preserve tumour shape, size, and intensity to ensure diagnostic reliability. Conventional GAN frameworks often prioritise overall visual realism while overlooking fine-grained lesion details. In brain tumour imaging, even minor distortions of tumour boundaries can degrade segmentation accuracy and potentially affect treatment planning \cite{ref32,ref34}. To address this issue, Xin et al. \cite{ref32} proposed TC-MGAN, which aligns tumour regions across modalities to improve lesion consistency; however, it focuses primarily on tumour areas and neglects other diagnostically relevant structures.
	
	In this paper, we propose 3D-MC-SAGAN (Multi-Contrast Self-Attention GAN), a framework designed to overcome limitations of existing MRI modality synthesis approaches and generate high-quality volumes through attention-augmented learning. The model introduces a memory-bounded hybrid attention block that captures long-range spatial dependencies while maintaining computational efficiency for 3D volumes. Inspired by the success of attention mechanisms in deep learning, the framework dynamically focuses on salient image regions, enabling selective emphasis on clinically relevant structures and improving synthesis reliability in medical imaging tasks \cite{ref35}. The proposed 3D-MC-SAGAN performs multi-contrast MRI synthesis from a single input modality by conditioning a shared generator-critic pair on a contrast code within a unified architecture. The model employs a fully 3D encoder--decoder with residual connections and attention modules to capture both global anatomical context and fine-grained details while preserving tumour characteristics. To further guide synthesis, we incorporate a pre-trained 3D U-Net segmenter and optimise a composite objective combining adversarial, reconstruction, perceptual, and structural similarity losses with an explicit tumour-consistency constraint. Together, these components address key limitations of prior methods, including one-to-one translation, 2D or patch-based processing, contrast-agnostic conditioning, and purely intensity-driven objectives, enabling anatomically coherent and tumour-preserving synthesis across all target contrasts within a unified model. Our code is available at https://github.com/Z-202/mc-sagan.

	To summarise, our main contributions are as follows:
	\begin{itemize}
		
		\item We propose 3D-MC-SAGAN, a unified 3D conditional WGAN-GP framework for multi-contrast MRI synthesis. Conditioned on a T2w volume and a target-contrast code, the model generates T2f, T1c, or T1n within a single architecture, eliminating the need for separate modality-specific networks and reducing both training and deployment complexity. A conditional PatchGAN critic with an auxiliary domain-classification head further enforces contrast-consistent synthesis. 
		\item We introduce a Memory-Bounded Hybrid Attention (MBHA) block, designed for efficient attention in high-resolution 3D volumes. MBHA adaptively downsamples query and key-value tensors to keep attention computations within fixed token and matrix-size budgets, ensuring predictable GPU memory usage. When attention becomes infeasible, the block falls back to a lightweight squeeze-and-excitation mechanism, maintaining contextual modulation while avoiding excessive computational cost. 
		\item We design a multi-scale 3D encoder--decoder generator integrating residual blocks, attention-gated skip connections, standard non-local self-attention, and the proposed MBHA module. This architecture improves global-local feature aggregation while preserving fine-grained anatomical details, particularly in tumour regions.
		\item We incorporate tumour-aware learning objectives to encourage pathology-preserving synthesis. A frozen 3D U-Net provides segmentation guidance through a tumour-consistency loss combining BCE-with-logits and Dice terms, while MS-SSIM and a masked 3D perceptual loss based on MedicalNet features enhance perceptual fidelity and structural preservation. Extensive experiments on BraTS2023 demonstrate consistent improvements over strong state-of-the-art baselines. 
	\end{itemize}

	\section{Related Work}
	
	Early deep learning approaches to medical image synthesis often relied on patch-based or locally constrained convolutional models trained with pixel-wise reconstruction losses \cite{ref27,ref36,ref37,ref38,ref39,ref40,ref41}. Operating on limited spatial neighbourhoods, these methods lacked the capacity to model long-range dependencies and global anatomical structure. To address this limitation, whole-image encoder--decoder architectures, such as fully convolutional networks and U-Nets, were developed to capture broader contextual information.
	However, despite this architectural shift, optimisation was still typically driven by per-pixel $\ell_1$ or $\ell_2$ losses, which tend to favour local intensity accuracy \cite{ref42,ref43,ref44}. A well-documented drawback of purely pixel-aligned loss functions is the tendency to produce blurry outputs that lack high-frequency detail. In essence, minimising mean squared error encourages an average of plausible textures, yielding high PSNR but over-smoothed images that are not perceptually sharp. This limitation motivated the adoption of adversarial training and more sophisticated objectives to recover realistic details \cite{ref30,ref45,ref46}.
	
	Generative Adversarial Networks (GANs) \cite{goodfellow2014gan} introduced an adversarial framework in which a discriminator guides the generator to approximate the target data distribution. Conditional GANs (cGANs) \cite{mirza2014cgan} extend this idea by conditioning both generator and discriminator on an observed input $x$, enabling supervised mappings $x \rightarrow y$ rather than unconditional sampling. In image-to-image translation, adversarial loss is commonly combined with a pixel-wise term (e.g., $\ell_1$) to preserve structural correspondence while improving perceptual fidelity \cite{ref30}. For example, pix2pix \cite{ref30} demonstrates that augmenting reconstruction loss with adversarial training reduces over-smoothing and restores sharper high-frequency detail compared with reconstruction-only optimisation.
	
	Unpaired translation frameworks, such as CycleGAN \cite{ref31}, further relaxed the requirement for paired training data by introducing cycle-consistency constraints alongside distribution alignment. However, methods that primarily enforce target-domain distribution matching may introduce bias, potentially distorting source-specific structures or subtle anatomical details.
	If certain clinical features are under- or over-represented in the target distribution, the model may introduce or suppress structures (e.g., tumours) to better match the distribution. Cycle-consistency alone does not guarantee a correct mapping, so features can still be hallucinated or improperly encoded, and translated images may fail to preserve clinically relevant labels, posing a risk for misdiagnosis. Thus, while adversarial losses are essential for restoring high-frequency detail, they are often combined with additional constraints, such as identity, cycle-consistency, or content losses, to preserve anatomy in medical image synthesis \cite{ref47}.
	
	A prominent line of research incorporates edge- or structure-aware objectives to preserve anatomical boundaries and textural fidelity. Ea-GANs \cite{ref19}, for example, extract Sobel-based edge maps and integrate them into both generator and discriminator, encouraging boundary-consistent synthesis. By enabling the discriminator to exploit explicit edge cues, the generator is steered toward sharper structural boundaries. MT-Net \cite{mtnet} extends this idea by replacing handcrafted edge operators with a self-supervised Edge-MAE that learns structural and contextual representations from both paired and unpaired data, thereby enhancing robustness and representation quality.
	To further improve anatomical consistency, subsequent work transitioned from 2D slice-wise synthesis to 3D volumetric and multi-resolution frameworks \cite{ref19,ref45,ref48,ref49,ref50,ref51}. slice-wise (2D) models are prone to inter-slice discontinuities when volumes are reconstructed, whereas 3D cGANs explicitly enforce coherence across all axes \cite{ref19}. Multi-Resolution Guided 3D GANs employ multi-resolution U-Nets on both generator and discriminator, together with voxel-wise losses at multiple scales, to sharpen fine-grained detail while stabilising training \cite{ref50}.  Comparative analyses indicate that off-plane discontinuities are more pronounced in 2D pix2pix than in 3D cGANs, supporting the transition to volumetric synthesis \cite{ref19}. Additionally, 3D patch-based formulations have demonstrated improved robustness to anatomical outliers \cite{ref52}.

	More recently, synthesis networks have increasingly adopted hybrid convolutional neural network (CNN)-transformer architectures \cite{ref20,luo2021transgan_pet}. In these designs, convolutional layers preserve local spatial precision, while self-attention mechanisms model long-range contextual dependencies. A common design pattern places transformers at lower spatial resolution inside an encoder--decoder framework, while retaining convolutional pathways to support fine-detail reconstruction. 
	Transformer-GAN follows this paradigm by embedding a transformer bottleneck within a convolutional encoder--decoder \cite{luo2021transgan_pet}, while the Fully Convolutional Transformer (FCT) demonstrates how transformer-style blocks can be integrated into skip-connected encoder--decoders to improve long-distance feature aggregation \cite{ref53}. Following this line of work, ResViT introduces aggregated residual transformer (ART) blocks in the bottleneck, combining transformer and residual CNN modules through residual connections to jointly model global context and localized representations \cite{ref20}.

	Alongside hybrids, transformer-centric synthesisers explore convolution-free or transformer-heavy pipelines and their associated trade-offs. PTNet3D addresses 3D MRI synthesis with pyramid transformers, demonstrating that transformer layers bring strong representational capacity but entail high computational and memory costs for volumetric data; consequently, designs prioritise parameter efficiency and careful bottlenecking \cite{ref51}. More generally, the self-attention mechanism offers long-range modelling advantages, motivating selective attention use or hybridisation \cite{ref51,ref20}. Where applicable, adversarial components (e.g., 3D patch-level discriminators) and hybrid losses (pixel-wise and perceptual) are used to stabilise optimisation and enhance realism \cite{ref51}.

	In cross-modality brain MRI synthesis, several representative models highlight remaining limitations. Ea-GANs \cite{ref19} introduce a 3D edge-aware cGAN trained on large overlapping $128\times128\times128$ patches from BraTS volumes. However, they are still restricted to single-contrast one-to-one mappings and rely on low-level edge cues without explicit tumour semantics or cross-contrast consistency constraints, and thus can still exhibit residual blurring of lesion interiors under a simple L1 and edge objective. MT-Net \cite{mtnet} proposes a two-stage, slice-wise transformer framework with Edge-MAE pretraining and dual-scale Swin-Transformer refinement, optimised using $\ell_1$ and feature-consistency losses but without adversarial supervision.

	ResViT \cite{ref20}, although combining convolutional encoders/decoders with a transformer bottleneck, operates in a 2D slice-wise manner and is mainly configured for one-to-one or many-to-one mappings without explicit tumour-aware constraints. PTNet3D \cite{ref51} employs a 3D pyramid transformer trained on fixed $64^3$ patches with GAN, $\ell_2$, and perceptual losses, yet remains patch-limited and strictly one-to-one in tumour-bearing settings, lacking mechanisms to preserve pathological anatomy. Multi-resolution 3D-mADUNet \cite{ref50} uses a dense-attention 3D generator with a multi-resolution discriminator and a 2.5D VGG-19 perceptual loss, but requires separate networks for each contrast pair and inherits limited global context and potential inter-patch inconsistencies from patch-wise inference. Diffusion-based approaches, such as cWDM \cite{cwdm}, implement 3D wavelet-domain diffusion with a 3D U-Net denoiser operating on full-resolution volumes. However, synthesis is realised via multiple independent three-to-one models trained with an MSE objective in wavelet space, limiting cross-contrast structural sharing and often yielding over-smoothed solutions. Moreover, tumour-specific or clinically salient structures are not explicitly constrained.

	\section{Methodology}
\begin{figure*}[t]
		\centering
		\includegraphics[width=0.9\textwidth]{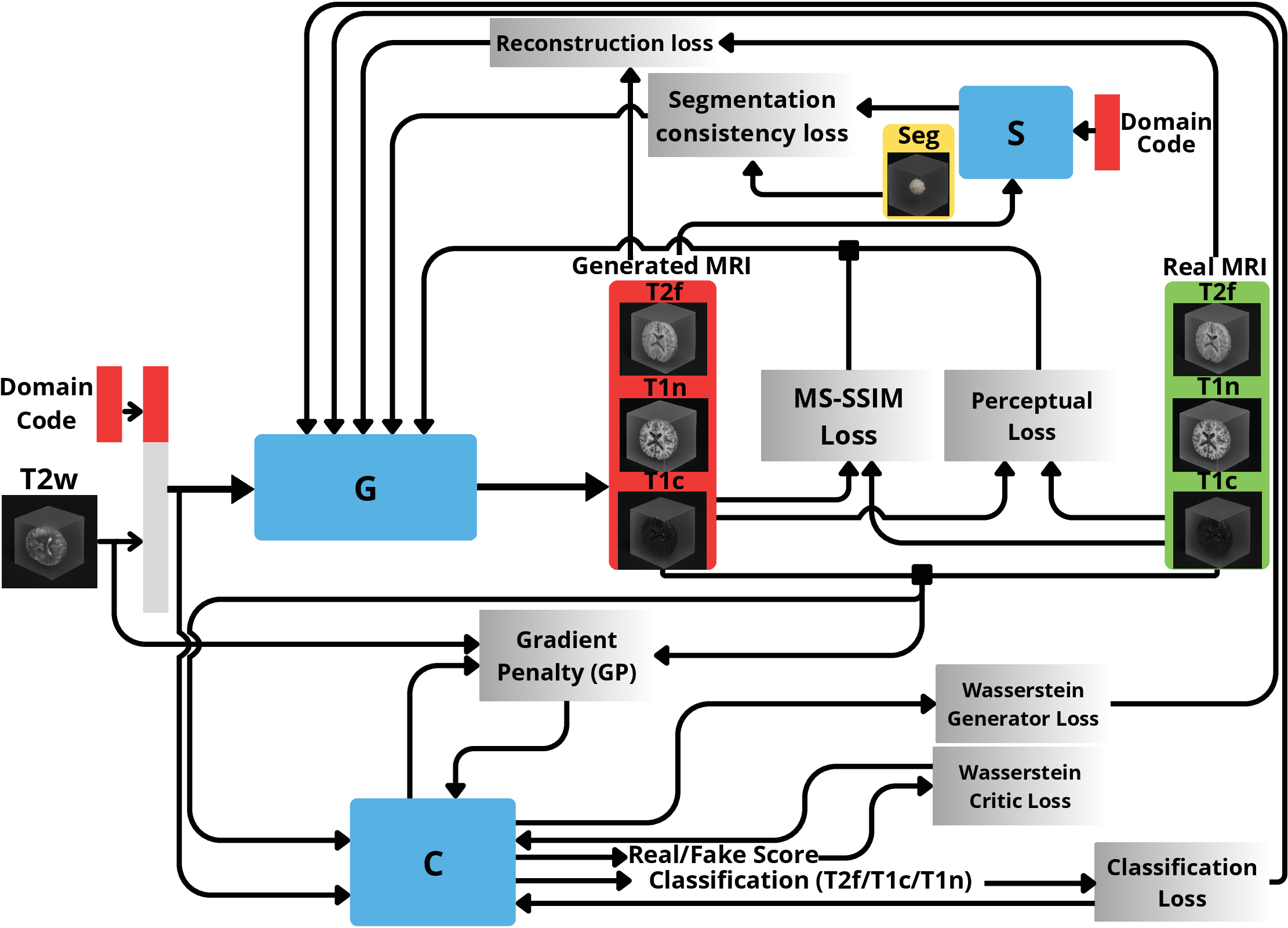}
		\caption{Overview of 3D-MC-SAGAN framework.}
		\label{fig1}
	\end{figure*}

	The proposed 3D-MC-SAGAN jointly optimises two objectives. A single conditional model synthesises the T2f, T1c, and T1n modalities from a T2w input, where a one-hot domain code $c \in \{0,1\}^3$ selects the target contrast. In parallel, the framework improves tumour segmentation performance in the synthesised domain. To support these objectives, the overall system is composed of four core components: the critic, the segmenter, the generator, and the memory-bounded hybrid attention block. Fig.\ref{fig1} illustrates the overall architecture of the proposed framework and interactions between these components.
	
	\subsection{Critic $\mathcal{C}$}
	
	The critic $\mathcal{C}$ serves a dual role: it estimates the Wasserstein distance between real and generated target volumes and performs auxiliary domain classification over the three target contrasts ${\text{T2f}, \text{T1c}, \text{T1n}}$. 
	Given a source T2w volume $x_s$ and a domain label $c$, the generator produces a target prediction $\hat{y}_c = G(x_s, c)$. For each training sample, the critic receives the concatenation of a target volume, either real ($y_c$) or generated ($\hat{y}_c$), with the source volume $x_s$. It outputs a 3D patch realism map together with a three-way class logit vector. The gradient penalty is computed from the spatial sum of the patch realism map.
	Under the WGAN-GP formulation~\cite{wgangp}, the adversarial loss for the critic is
	\begin{equation}
		\mathcal{L}_{\mathcal{C}}^{\mathrm{WGAN}} 
		= \mathbb{E}_{x_s, y_c}\!\left[\,\mathcal{C}_{\mathrm{adv}}(\hat{y}_c, x_s)\right]
		- \mathbb{E}_{x_s, y_c}\!\left[\,\mathcal{C}_{\mathrm{adv}}(y_c, x_s)\right],
	\end{equation}
	where $\mathcal{C}_{\mathrm{adv}}(\cdot,\cdot)$ denotes the spatially averaged realism score, $y_c$ is the real target volume in contrast $c$, and $\hat{y}_c = G(x_s, c)$ is the corresponding synthetic volume. Minimizing this loss encourages higher scores for real targets than for generated targets.
	To enforce the $1$-Lipschitz constraint, we regularize the critic with a gradient penalty term:
	\begin{equation}
		\mathcal{L}_{\mathrm{GP}} 
		= \mathbb{E}_{x_s, y_c, \hat{y}_c, \alpha}\!\left[\Big(\big\|\nabla_{\tilde{y}_c}\,\mathcal{C}_{\mathrm{adv}}(\tilde{y}_c, x_s)\big\|_2 - 1\Big)^2\right],
	\end{equation}
	where $\alpha \sim \mathcal{U}[0,1]$ and 
	$\tilde{y}_c = \alpha\, y_c + (1-\alpha)\,\hat{y}_c$ is a random interpolation between real and generated targets. In the implementation, this penalty is scaled by a factor $\lambda_{\mathrm{GP}}$.
	
	The critic also learns to classify each real target volume into its correct contrast. Let $\mathcal{C}_{\mathrm{cls}}(c \mid y_c, x_s)$ denote the softmax probability assigned to class $c$ by the classification head of $\mathcal{C}$ when fed the pair $(y_c, x_s)$. The domain classification loss for the critic is:
	\begin{equation}
		\mathcal{L}_{\mathrm{cls}}^{\mathcal{C}}
		=
		\mathbb{E}_{x_s,c}
		\big[
		\ell_{\mathrm{CE}}\big(\mathcal{C}_{\mathrm{cls}}(y_c,x_s),\,c\big)
		\big],
	\end{equation}
	where $\mathcal{C}_{\mathrm{cls}}(\cdot,\cdot)$ denotes the classification logits and $\ell_{\mathrm{CE}}$ is the cross-entropy loss.

	The overall objective for the critic in 3D-MC-SAGAN is:
	\begin{equation}
		\mathcal{L}_{\mathcal{C}} 
		= \mathcal{L}_{\mathcal{C}}^{\mathrm{WGAN}}
		+ \lambda_{\mathrm{GP}}\,\mathcal{L}_{\mathrm{GP}}
		+ \lambda_{\mathrm{cls}}\,\mathcal{L}_{\mathrm{cls}}^{\mathcal{C}},
	\end{equation}
	where $\lambda_{\mathrm{GP}}$ and $\lambda_{\mathrm{cls}}$ control the strength of the gradient penalty and domain classification terms, respectively.

	\subsection{Segmenter $\mathcal{S}$}
	We employ a conditional 3D U-Net $\mathcal{S}$ as a tumour segmenter within the 3D-MC-SAGAN framework. 
	This network is pre-trained on a standalone tumour segmentation task. 
	For each training sample, one of the three target contrasts $\{\text{T2f}, \text{T1c}, \text{T1n}\}$ is randomly selected as input with the corresponding contrast code, and the network is trained to predict the corresponding tumour mask. The segmenter is optimised with Dice and binary cross-entropy losses:
	\begin{equation}
		\small
		\mathcal{L}_{\mathcal{S}}=\mathcal{L}_{\mathrm{Dice}}+0.5\mathcal{L}_{\mathrm{BCE}},\;
		\mathcal{L}_{\mathrm{Dice}}=1-\frac{2\sum_i \hat{m}(i)\mathrm{gt}(i)+\epsilon}
		{\sum_i \hat{m}(i)+\sum_i \mathrm{gt}(i)+\epsilon}.
	\end{equation}
	where $\hat{m}$ is the predicted probability map, {gt} is the ground truth mask, $\mathcal{L}_{\mathrm{BCE}}$ is applied to the logits, and $\epsilon=10^{-6}$.
	
	After pre-training, we load the learned parameters into $\mathcal{S}$ and keep all weights frozen during GAN training. 
	In the 3D-MC-SAGAN stage, $\mathcal{S}$ is applied to the generated target volumes and its output is compared with the ground-truth tumour masks with an equally weighted Dice-BCE loss. 
	Thus, $\mathcal{S}$ acts as a fixed segmentation backbone that imposes a tumour-aware constraint on the generator and helps preserve lesion structure in the synthesised volumes.

	\subsection{Generator $\mathbf{G}$}
\begin{figure*}[t]
	\centering
	\includegraphics[width=\textwidth]{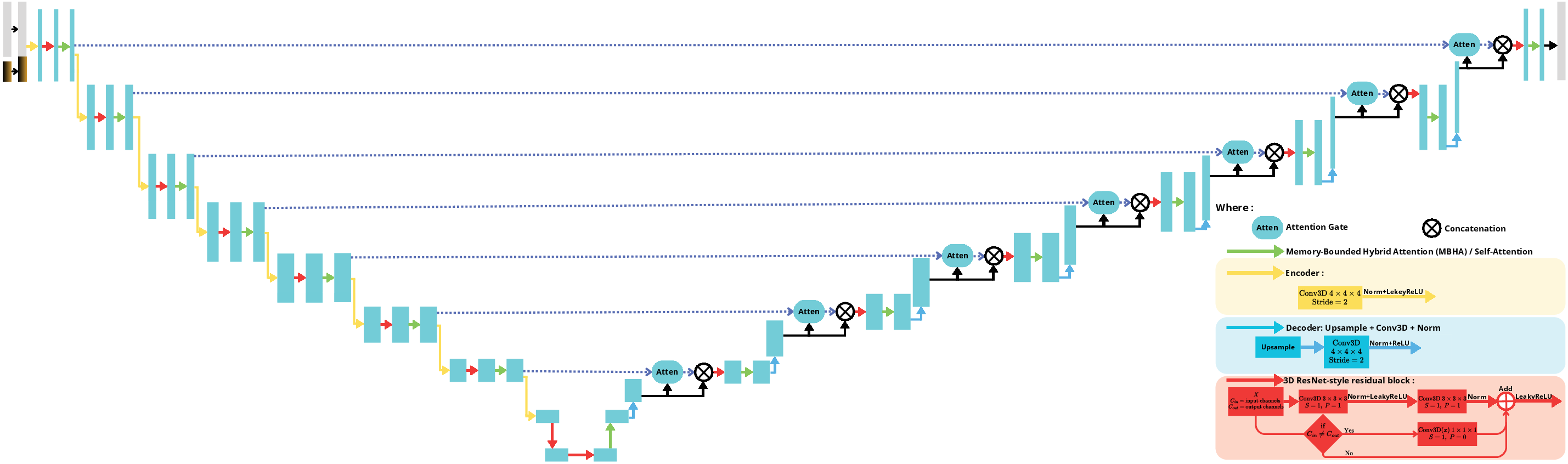}
	\caption{Generator architecture. A 3D encoder-decoder with residual blocks, attention-gated skip connections, and mixed attention.}
	\label{fig:fig2}
\end{figure*}

	The Generator $\mathbf{G}$ is implemented as a 3D attention U-Net with residual blocks and domain conditioning, as illustrated in Fig.~\ref{fig:fig2}. It maps a T2w input volume and a contrast code to a synthesised target contrast. The network operates on full 3D volumes rather than slices. This design preserves through-plane consistency and volumetric tumour morphology during synthesis. Its overall structure follows encoder--decoder generators used in cross-modality MR synthesis, but extends them with volumetric attention, attention-gated skips, and multi-domain conditioning.

	Given a source T2w volume $x_s \in \mathbb{R}^{1 \times D \times H \times W}$ and a one-hot domain code $c \in \{0,1\}^3$, the generator first reshapes $c$ to $(3, D, H, W)$ and concatenates it with $x_s$ along with the channel dimension to form a four-channel input $x_{\text{in}} \in \mathbb{R}^{4 \times D \times H \times W}$. The contrast code selects the desired target domain within a single shared generator. The network outputs a single-channel synthetic target volume $\hat{y}_k = \mathbf{G}(x_s, c) \in [-1,1]^{1 \times D \times H \times W}$ for a domain index $k \in \{0,1,2\}$ (T2f, T1c, or T1n).

	\noindent\textbf{Encoder:}
	The encoder path consists of a sequence of 3D convolutional layers followed by residual blocks and attention modules, including the MBHA block described in Section~\ref{sec:mbha}. Each level reduces spatial resolution and increases the number of feature channels. Instance normalisation and leakyReLU activations are applied after each downsampling convolution except at the final bottleneck. The residual blocks refine features within each scale. They stabilise training and improve gradient flow through the deep 3D stack. In the first three encoder stages, MBHA is applied to the relatively large 3D feature maps to preserve global context while keeping memory usage bounded. In the subsequent deeper attention-equipped stages, where the spatial size is smaller, the encoder instead employs full 3D self-attention to model richer long-range interactions at manageable computational cost. The overall design enables the encoder to capture both local texture and larger-scale anatomical context. It also enhances long-range contextual interactions between distant tumour and tissue regions.
	
	\noindent\textbf{Bottleneck:}
	The bottleneck operates on the most compressed volumetric representation. It receives the deepest encoder features and applies additional residual refinement. Group Normalisation replaces Instance Normalisation at this point. This prevents degenerate statistics when the spatial support collapses to a very small number of voxels. A 3D self-attention is then applied at the bottleneck. This step enriches the latent code with long-range interactions before decoding.
	
	\noindent\textbf{Decoder:}
	The decoder mirrors the encoder with a sequence of 3D upsampling convolutions that progressively upsample the feature maps back to the original resolution. At each scale, an attention gate modulates the corresponding encoder skip connection using the current decoder features as a gating signal. The gate produces a voxel-wise attention map over the encoder features. This mechanism reduces the influence of less relevant features and focuses the network on spatial regions that are most informative for the synthesis task. The gated encoder features are concatenated with the upsampled decoder features. A residual block then fuses these streams and refines the combined representation. At lower resolutions, the decoder employs full 3D self-attention, whereas at higher resolutions it uses MBHA to retain broader context under a bounded memory budget.

	\noindent\textbf{Generator Loss Functions:} We train $G$ using a weighted combination of adversarial, classification, reconstruction, perceptual, SSIM, and segmentation losses.

	\noindent\textbf{Adversarial loss ($\mathcal{L}_{\text{adv}}$):}
	An adversarial term is used to match the distribution of synthesised 3D target volumes to that of real targets conditioned on the same T2w source. The generator is trained to increase this score for its outputs. The adversarial loss for the generator follows the WGAN formulation and uses the critic output on fake samples:
	\begin{equation}
		\mathcal{L}_{G}^{\mathrm{adv}}
		=
		- \mathbb{E}_{x_s, c}
		\big[
		\mathcal{C}_{\mathrm{adv}}
		\big(
		\hat{y}_c, x_s
		\big)
		\big]
	\end{equation}

	\noindent\textbf{Domain classification loss ($\mathcal{L}_{\text{cls}}$):}
	A domain classification loss encourages the generator to produce a volume whose contrast matches the requested label:
	\begin{equation}
		\mathcal{L}_{\mathrm{cls}}^{G}
		=
		\mathbb{E}_{x_s, c}
		\big[
		\ell_{\mathrm{CE}}
		\big(
		\mathcal{C}_{\mathrm{cls}}(\hat{y}_c, x_s),\, c
		\big)
		\big]
	\end{equation}

	where $\ell_{\mathrm{CE}}$ denotes the cross-entropy loss.

	\noindent\textbf{Reconstruction loss ($\mathcal{L}_{\mathrm{rec}}$):}
	To promote faithful cross-modality synthesis while emphasising clinically relevant regions, we use a \emph{segmentation-weighted} voxel-wise $\ell_1$ reconstruction loss,  assigning higher penalties to voxels inside the tumour mask than to background voxels. Let $m \in {0,1}^{H\times W\times D}$ denote the binary tumour mask derived from the ground-truth segmentation, and $\alpha=4$ controls the degree of emphasis. A spatial weight map is defined as $w = 1 + \alpha m$, and the weighted reconstruction objective is computed as:
	\begin{equation}
		\mathcal{L}_{\mathrm{rec}}
		=
		\mathbb{E}_{x_s, c}
		\left[
		\frac{1}{|\Omega|}
		\sum_{v \in \Omega}
		w(v)\, \big| y_c(v) - \hat{y}_c(v) \big|
		\right],
	\end{equation}
	where $\Omega$ indexes all voxels. This formulation retains anatomical fidelity while encouraging accurate reconstruction within tumour regions, consistent with the weighted $\ell_1$ implementation used during training.
	
	\noindent\textbf{Perceptual loss ($\mathcal{L}_{\mathrm{perc}}$):}
	Voxel-wise objectives are insufficient for enforcing high-level anatomical structure and modality-specific contrast.
	A 3D perceptual constraint is therefore applied in a deep feature space extracted from a fixed 3D MedicalNet ResNet-50 backbone.
	Feature activations are taken from four residual stages, namely \texttt{layer1}, \texttt{layer2}, \texttt{layer3}, and \texttt{layer4}.
	An $\ell_{1}$ distance is computed between the generated and real feature maps.
	Layer-wise weighting is used to balance low-level detail and higher-level structural cues.
	The corresponding weights are set to $(\lambda_1,\lambda_2,\lambda_3,\lambda_4)=(1.0,0.5,0.25,0.1)$. 
	The perceptual loss is then computed as follows:
	
	\begin{equation}
		\mathcal{L}_{\mathrm{perc}}
		=
		\mathbb{E}_{x_s,c}
		\left[
		\sum_{k=1}^{4}
		\lambda_k\,
		\left\|
		\Phi_k\!\left(y_c\right)
		-
		\Phi_k\!\left(\hat{y}_c\right)
		\right\|_{1}
		\right]
	\end{equation}
	
	\noindent
	where $\Phi_k(\cdot)$ denotes the feature tensor extracted at stage $k$ of the MedicalNet backbone.
	The $\ell_{1}$ norm is evaluated as the mean absolute difference over all feature elements.
	This term encourages anatomically plausible synthesis and preserves contrast characteristics that are not fully captured by reconstruction losses.

	\noindent\textbf{MS-SSIM loss ($\mathcal{L}_{\text{MS-SSIM}}$):}
	Structural similarity is enforced through a multi-scale SSIM loss. The loss is defined as:
	\begin{equation}
		\mathcal{L}_{\mathrm{MS\text{-}SSIM}}
		=
		\mathbb{E}_{x_s, c}
		\big[
		1 - \mathrm{MS\text{-}SSIM}
		\big(
		y_c, \hat{y}_c
		\big)
		\big].
	\end{equation}
	This term encourages the synthesised volumes to preserve local structural patterns and contrast relationships that are not fully constrained by voxel-wise reconstruction.

	\noindent\textbf{Segmentation-consistency loss ($\mathcal{L}_{\mathrm{seg}}$):}
	A segmentation-consistency term is introduced to preserve pathology during cross-modality synthesis.
	A pretrained conditional 3D segmentation network $S$ is kept frozen and used as an auxiliary supervisor.
	The generated volume $\hat{y}_c$ and the contrast code $c$ are forwarded through $S$ to obtain voxel-wise logits
	$\ell = S(\hat{y}_c, c)$ and probabilities $\hat{m}=\sigma(\ell)$, where $\sigma(\cdot)$ denotes the sigmoid function.
	Supervision is provided by the binary tumour mask $m$.
	The segmentation-consistency objective is defined as an equal-weight combination of a voxel-wise binary cross-entropy loss on logits and a soft Dice loss:
	\begin{equation}
		\begin{aligned}
			\mathcal{L}_{\mathrm{seg}}
			&=
			\mathbb{E}_{x_s,c}
			\Big[
			\tfrac{1}{2}\,\mathcal{L}_{\mathrm{BCE}}(\ell, m)
			+
			\tfrac{1}{2}\,\mathcal{L}_{\mathrm{Dice}}(\hat{m}, m)
			\Big], \\
			&\qquad
			\ell = S(\hat{y}_c,c), \;\; \hat{m}=\sigma(\ell).
		\end{aligned}
	\end{equation}

	The overall generator objective is a weighted sum of the individual loss terms:
	\begin{equation}
		\begin{split}
			\mathcal{L}_{G}
			=
			\mathcal{L}_{G}^{\mathrm{adv}}
			+ \lambda_{\mathrm{cls}} \mathcal{L}_{\mathrm{cls}}^{G}
			+ \lambda_{\mathrm{rec}} \mathcal{L}_{\mathrm{rec}} \\
			\quad
			+ \lambda_{\mathrm{seg}} \mathcal{L}_{\mathrm{seg}}
			+ \lambda_{\mathrm{perc}} \mathcal{L}_{\mathrm{perc}}
			+ \lambda_{MS\text{-}SSIM} \mathcal{L}_{MS\text{-}SSIM}.
		\end{split}
	\end{equation}

	\subsection{Memory-Bounded Hybrid Attention Block}\label{sec:mbha}
\begin{figure*}[t]
	\centering
	\includegraphics[width=0.9\textwidth]{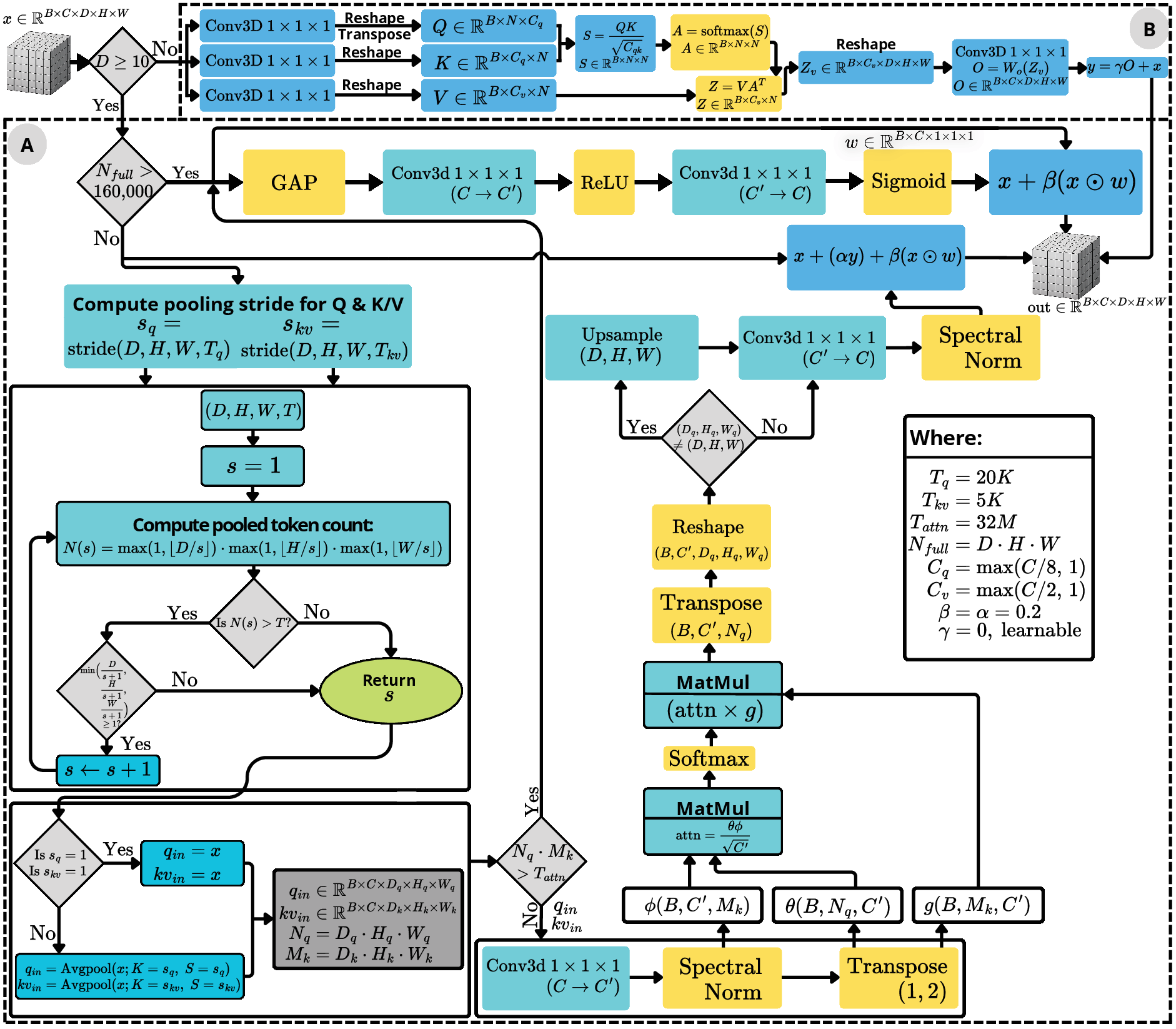}
	\caption{Attention mechanisms used in the generator: (A) MBHA for high-resolution stages and (B) standard 3D self-attention for low-resolution stages.}
	\label{fig:fig__2}
\end{figure*}

	The proposed Memory-Bounded Hybrid Attention (MBHA) block implements a 3D self-attention module that fuses channel recalibration and non-local spatial attention under an explicit memory budget. It is built in two established attention paradigms: SE-based channel gating \cite{Hu_2018_CVPR} and embedded-Gaussian non-local operators with tokenised spatial dimensions \cite{Wang_2018_CVPR}. The block is conceptually related to global-context modelling \cite{Cao_2019_ICCV}, self-attention GANs \cite{pmlr-v97-zhang19d}, and asymmetric non-local designs \cite{Zhu_2019_ICCV}. 
	While prior approaches - such as GCNet’s query-independent global context vectors on 2D feature maps \cite{Cao_2019_ICCV}, non-local GAN blocks without attention-size control \cite{pmlr-v97-zhang19d}, and asymmetric pooling strategies to reduce computational cost \cite{Zhu_2019_ICCV} - have explored combinations of channel and spatial attention, they are typically limited to 2D formulations and lack explicit mechanisms for memory regulation. In contrast, MBHA unifies and extends these ideas within a single volumetric module that enforces predictable, resolution-robust memory usage, thereby preventing out-of-memory failures in high-resolution 3D MRI synthesis while still enabling adaptive long-range attention when computational budgets permit. Figure \ref{fig:fig__2} illustrates the overall architecture of MBHA.
	
	Given an input feature map $x \in \mathbb{R}^{B\times C\times D\times H\times W}$ (batch $B$, channels $C$, spatial dimensions $D\times H\times W$), the block computes an output $y$ of the same shape via two branches: a squeeze-and-excitation (SE) channel attention branch~\cite{Hu_2018_CVPR} and a pooled self-attention branch operating on adaptively downsampled feature maps. Both branches are combined in a residual fashion.
	\\
	\textbf{SE channel gating.}
	The SE branch follows the formulation of Hu \emph{et al.}~\cite{Hu_2018_CVPR}.
	A global channel descriptor is obtained by adaptive average pooling:
	\[
	z = \mathrm{GAP}(x)\in \mathbb{R}^{B\times C\times 1\times 1\times 1},
	\]
	which then passes through a bottleneck of $1\times1\times1$ convolutions with ReLU and sigmoid to generate channel-wise modulation weights as follows:,
	\[
	s(x) = \sigma\!\big(W_2\,\delta(W_1 z)\big)\in \mathbb{R}^{B\times C\times 1\times1\times1},
	\]
	where $\delta(\cdot)$ denotes ReLU and $\sigma(\cdot)$ denotes the sigmoid.
	The resulting gate rescales the input channels via broadcasting,
	$
	x_{\text{SE}} = x \odot s(x),
	$
	with $\odot$ denoting channel-wise multiplication.
	The SE bottleneck uses an intermediate width $C'=\max(C/r,8)$ with reduction ratio $r=8$, and this reduced dimension is shared with the non-local branch.
	\\
	\textbf{Memory-limited self-attention.}
	Direct 3D non-local attention on $x$ requires an affinity matrix of size $N\times N$ with $N=DHW$, which is prohibitive at volumetric resolutions~\cite{Wang_2018_CVPR,Zhu_2019_ICCV,Cao_2019_ICCV}.
	To bound the memory footprint, the block introduces three caps.
	These are $T_q$ (maximum number of query tokens), $T_{kv}$ (maximum number of key/value tokens) and $T_{\text{attn}}$ (maximum number of attention elements $N_qM_k$).
	We first compute the total number of voxels as $N_{\text{full}} = D H W$.
	If $N_{\text{full}}$ exceeds a constant multiple of the query budget (in practice $N_{\text{full}} > \kappa T_q$ with $\kappa=8$), the block activates a fail-safe mode and bypasses the non-local branch.
	In that case the attention reduces to SE-only, described later in this section.
	
	When $N_{\text{full}}$ is within this range, the module derives separate downsampling strides for queries and keys/values.
	For a given budget $T$, the stride $s$ is chosen as the smallest integer $s\ge 1$ such that the token count after strided average pooling does not exceed $T$ and at least one voxel remains along each spatial axis.
	Formally, the query and key/value strides are:
	\[
	s_q \;=\; \min\Big\{s\ge1:\;\Big\lfloor\frac{D}{s}\Big\rfloor\;\Big\lfloor\frac{H}{s}\Big\rfloor\;\Big\lfloor\frac{W}{s}\Big\rfloor \le T_q\Big\},
	\]
	\[
	s_{kv} \;=\; \min\Big\{s\ge1:\;\Big\lfloor\frac{D}{s}\Big\rfloor\;\Big\lfloor\frac{H}{s}\Big\rfloor\;\Big\lfloor\frac{W}{s}\Big\rfloor \le T_{kv}\Big\}.
	\]
	Average pooling with kernel size and stride equal to $s_q$ or $s_{kv}$ gives pooled tensors:
	\[
	x_q = P_{s_q}(x), \qquad x_{kv} = P_{s_{kv}}(x).
	\]
	Let $D_q\times H_q\times W_q$ and $D_k\times H_k\times W_k$ denote the spatial sizes of $x_q$ and $x_{kv}$ respectively. We define $N_q$ and $M_k$ as follows:
	\[
	N_q = D_q H_q W_q,\qquad M_k = D_k H_k W_k.
	\]
	If the product $N_q M_k$ exceeds $T_{\text{attn}}$, the non-local computation is skipped and the block again falls back to SE-only.
	These two checks, on $N_{\text{full}}$ and $N_qM_k$ respectively, impose an explicit token and affinity budget, which is not present in previous non-local formulations~\cite{Wang_2018_CVPR,Zhu_2019_ICCV,Cao_2019_ICCV}.
	\\
	\textbf{Fail-safe SE-only mode.}
	When either budget constraint is violated, the output is computed as a purely channel-wise refinement.
	The block adds a scaled SE residual to the input,
	\[
	y = x + \beta\, x_{\text{SE}},
	\]
	with a small scaling factor $\beta=0.2$.
	This fail-safe mode keeps memory usage bounded for arbitrarily large volumetric inputs, while still exploiting SE-based global channel recalibration~\cite{Hu_2018_CVPR}.
	In practice, this mode is active at very high spatial resolutions, where full non-local attention would be most expensive~\cite{Wang_2018_CVPR,Zhu_2019_ICCV,Cao_2019_ICCV}.
	\\
	\textbf{Pooled non-local attention.}
	When the budgets are satisfied, the block computes a pooled non-local self-attention on $x_q$ and $x_{kv}$.
	Query, key and value tensors are respectively obtained by spectral-normalised $1\times1\times1$ convolutions,
	\[
	Q = \Theta(x_q),\quad K = \Phi(x_{kv}),\quad V = G(x_{kv}),
	\]
	where $\Theta$,$\Phi$, and $G$ map $C$ channels to $C'=\max(C/r,8)$ channels respectively.
	After flattening spatial dimensions, we obtain:
	\[
	Q \in \mathbb{R}^{B\times N_q\times C'},\quad
	K \in \mathbb{R}^{B\times C'\times M_k},\quad
	V \in \mathbb{R}^{B\times M_k\times C'}.
	\]
	A scaled dot-product affinity is then computed as
	\[
	A = \mathrm{softmax}\!\Big(\tfrac{QK}{\sqrt{C'}}\Big)\in\mathbb{R}^{B\times N_q\times M_k},
	\]
	with the softmax applied along the key dimension.
	This corresponds to the embedded-Gaussian instantiation of the non-local operator~\cite{Wang_2018_CVPR}, using a Transformer-style scaling and asymmetric tokens as in self-attention GANs and asymmetric non-local networks~\cite{pmlr-v97-zhang19d,Zhu_2019_ICCV}.
	The attended response is
	\[
	Y = A V \in\mathbb{R}^{B\times N_q\times C'},
	\]
	which is reshaped into a feature map of size $B\times C'\times D_q\times H_q\times W_q$.
	If pooling was applied ($s_q>1$), this tensor is upsampled back to $(D,H,W)$ by trilinear interpolation.
	A spectral-normalised $1\times1\times1$ convolution $W_O$ then projects this map to $C$ channels.
	We denote this non-local output as
	\[
	x_{\text{NL}} = W_O(\mathrm{Up}(Y)).
	\]
	Finally, the block returns
	\[
	y = x + \alpha\, x_{\text{NL}} + \beta\, x_{\text{SE}},
	\]
	with $\alpha=0.2$ and $\beta=0.2$ in our implementation.
	The residual scaling keeps the non-local correction small at the start of training and improves stability, following the spirit of self-attention GANs~\cite{pmlr-v97-zhang19d} while also controlling the SE contribution.

	\section{Experiments}
	
	\subsection{Data Preparation and Experimental Settings}
	The Brain Tumour Segmentation Challenge (BraTS 2023) dataset~\cite{brats} comprises 1251 subjects, each providing four co-registered MRI sequences and a tumour segmentation label with spatial dimension of $240 \times 240 \times 155$. Each subject includes T2w, T2f, T1n, T1c, and the corresponding tumour mask label volumes. T2w is used as the source modality to synthesise T2f, T1c, and T1n. All modalities are loaded as 3D NIfTI volumes. Each sequence is independently clipped to the $0.1$th-$99.9$th intensity percentile to suppress outliers and then rescaled to $[-1,1]$ using min-max normalisation. Each volume is embedded into a $256 \times 256 \times 160$ grid by padding the last 5 axial slices with the background value and applying symmetric in-plane padding. The tumour mask is binarised and padded to the same spatial size using zero-padding. 
	
	The BraTS 2023 training cohort ($n=1251$) was randomly partitioned at the subject level into four mutually disjoint subsets: 1061 subjects were used to train 3D-MC-SAGAN, 100 subjects were used to train the auxiliary 3D tumour segmentation network employed for downstream evaluation, 10 subjects were used to validate the segmentation stage, and 80 subjects were reserved for final testing. In addition, 70 subjects were randomly selected from the official BraTS 2023 validation cohort ($n=219$) and used as a separate validation set for monitoring, hyperparameter tuning, and model selection during 3D-MC-SAGAN development. All splits were fixed before experimentation, and the same training, validation, and test partitions were used for all compared synthesis models to ensure a fair comparison. Training is conducted for 210 epochs with a batch size of 2.  The generator and critic are optimised using Adam ($\beta_{1}=0.0$, $\beta_{2}=0.9$) with an initial learning rate of $1\times10^{-4}$, reduced by a factor of 0.5 at epochs 80, 140, and 200. The critic is updated three times per generator step, and a gradient penalty ($\lambda_{\mathrm{GP}}=10$) is applied on interpolated real and synthetic volumes to stabilise training.

	The generator is optimised with a weighted combination of adversarial, domain-classification, reconstruction, segmentation-consistency, perceptual, and MS-SSIM losses.
	Loss weights are selected via random search on the validation set: for each trial, candidate weights are randomly sampled and evaluated under the same training protocol.
	Configurations are ranked based on a validation criterion reflecting both synthesis fidelity (PSNR/SSIM/MSE) and downstream faithfulness (segmentation performance), and the best-performing set is fixed for all experiments. The selected weights are $\lambda_{\mathrm{rec}}=29.8016$, $\lambda_{\mathrm{MS\text{-}SSIM}}=13.7866$, $\lambda_{\mathrm{perc}}=1.7760$, $\lambda_{\mathrm{seg}}=1.0421$, $\lambda_{\mathrm{cls}}=0.4613$, and $\lambda_{\mathrm{GP}}=10$.

	\renewcommand{\arraystretch}{1}
	\begin{table*}[t]
		\centering
		\setlength{\tabcolsep}{2pt} 
		\caption{Quantitative evaluation results (PSNR$\uparrow$, SSIM$\uparrow$, MSE$\downarrow$) for T2f, T1n, and T1c. The best results are in \textbf{bold}.}
		\label{tab:table1}
		\resizebox{\textwidth}{!}{
			\begin{tabular}{|l|ccc|ccc|ccc|c|}
				\hline
				\multirow{2}{*}{\textbf{Method}} &
				\multicolumn{3}{c|}{\textbf{PSNR$\uparrow$}} &
				\multicolumn{3}{c|}{\textbf{SSIM$\uparrow$}} &
				\multicolumn{3}{c|}{\textbf{MSE$\downarrow$}} &
				\multirow{2}{*}{\textbf{2D/3D}} \\ \cline{2-10}
				& \textbf{T2f} & \textbf{T1n} & \textbf{T1c}
				& \textbf{T2f} & \textbf{T1n} & \textbf{T1c}
				& \textbf{T2f} & \textbf{T1n} & \textbf{T1c} 
				& {} \\ \hline
				\textbf{Pix2Pix}
				& 21.150$\pm$1.240 & 23.775$\pm$1.133 & 19.167$\pm$0.913
				& 0.862$\pm$0.0214 & 0.907$\pm$0.0240 & 0.846$\pm$0.018
				& 0.0080$\pm$0.0026 & 0.0043$\pm$0.0012 & 0.0123$\pm$0.0025
				& 2D \\ \hline
				\textbf{CycleGAN}
				& 20.903$\pm$1.892 & 22.055$\pm$1.268 & 19.435$\pm$1.506
				& 0.845$\pm$0.035 & 0.873$\pm$0.015 & 0.811$\pm$0.025
				& 0.0089$\pm$0.0043 & 0.0065$\pm$0.0024 & 0.0121$\pm$0.0053
				& 2D \\ \hline
				\textbf{3D-mADUNet}
				& 24.692$\pm$3.893 & 20.800$\pm$4.257 & 26.645$\pm$3.938
				& 0.913$\pm$0.030 & 0.919$\pm$0.040 & 0.935$\pm$0.025
				& 0.0049$\pm$0.0043 & 0.0124$\pm$0.0105 & 0.0033$\pm$0.0039
				& 3D \\ \hline
				\textbf{PTNet3D}
				& 21.267$\pm$0.942 & 21.824$\pm$0.691 & 21.746$\pm$1.050
				& 0.888$\pm$0.022 & 0.923$\pm$0.018 & 0.893$\pm$0.021
				& 0.0076$\pm$0.0019 & 0.0066$\pm$0.0011 & 0.0069$\pm$0.0022
				& 3D \\ \hline
				\textbf{Ea-GAN}
				& 24.671$\pm$2.381 & 26.001$\pm$1.812 & 23.175$\pm$1.614
				& 0.928$\pm$0.024 & 0.954$\pm$0.020 & 0.913$\pm$0.021
				& 0.0039$\pm$0.0025 & 0.0027$\pm$0.0014 & 0.0051$\pm$0.0021
				& 3D \\ \hline
				\textbf{MT-Net}
				& 24.602$\pm$2.325 & 25.164$\pm$2.263 & 25.271$\pm$2.911
				& 0.885$\pm$0.027 & 0.921$\pm$0.032 & 0.880$\pm$0.031
				& 0.0040$\pm$0.0024 & 0.0035$\pm$0.0021 & 0.0038$\pm$0.0032
				& 2D \\ \hline
				\textbf{ResViT}
				& 25.560$\pm$1.594 & 26.950$\pm$1.954 & 24.278$\pm$1.517
				& 0.927$\pm$0.022 & 0.945$\pm$0.022 & 0.910$\pm$0.023
				& 0.0029$\pm$0.0011 & 0.0022$\pm$0.0011 & 0.0039$\pm$0.0017
				& 2D \\ \hline
				\textbf{cWDM}
				& 26.466$\pm$2.184 & 27.868$\pm$2.617 & 25.651$\pm$2.270
				& 0.932$\pm$0.019 & 0.957$\pm$0.020 & 0.920$\pm$0.020
				& 0.0025$\pm$0.0014 & 0.0019$\pm$0.0014 & 0.0031$\pm$0.0018
				& 3D \\ \hline
				\textbf{TC-MGAN}
				& 25.627$\pm$1.662 & 26.648$\pm$2.143 & 26.164$\pm$2.470
				& 0.927$\pm$0.019 & 0.951$\pm$0.022 & 0.909$\pm$0.017
				& 0.0029$\pm$0.0012 & 0.0024$\pm$0.0013 & 0.0029$\pm$0.0021
				& 2D \\ \hline
				\textbf{3D-MC-SAGAN}
				& \textbf{27.680$\pm$2.104} & \textbf{28.428$\pm$2.218} & \textbf{26.738$\pm$2.573}
				& \textbf{0.947$\pm$0.014} & \textbf{0.966$\pm$0.013} & \textbf{0.943$\pm$0.015}
				& \textbf{0.0019$\pm$0.0011} & \textbf{0.0016$\pm$0.0009} & \textbf{0.0025$\pm$0.0017}
				& \textbf{3D} \\ \hline
		\end{tabular}}
	\end{table*}
	\subsection{Evaluation Metrics}
	We evaluate synthesis fidelity using peak signal-to-noise ratio (PSNR), structural similarity index (SSIM) \cite{ref54}, and mean square error (MSE) and compare the  proposed 3D-MC-SAGAN against state-of-the-art methods. To assess distributional similarity, we compute a Fr\'echet-distance metric in the feature space of a 3D ResNet-50 backbone from MedicalNet, pre-trained on large-scale medical imaging datasets. This metric, referred to as MedicalNet Fr\'echet Distance (MFD), follows the FID principle but replaces the Inception network with a domain-specific 3D encoder\cite{refFID,refMed3D}. For downstream evaluation, tumour segmentation performance is measured using the Dice score by concatenating the synthesised modalities (T1n, T1c, and T2f) with the original T2w input and feeding them to the segmentation network.

	\begin{table}[h]
		\centering
		\small
		\setlength{\tabcolsep}{3.5pt}
		\renewcommand{\arraystretch}{1}
		\caption{{MFD results ($\times 10^{3}$) for synthesised MRI modalities (T2f, T1, and T1c). Lower is better ($\downarrow$).}}
		\label{tab:mfd_results}
		\begin{tabular}{|c|c|c|c|}
			\hline
			\textbf{Method} & \textbf{MFD $\downarrow$ (T2f)} & \textbf{MFD $\downarrow$ (T1n)} & \textbf{MFD $\downarrow$ (T1c)} \\
			\hline
			\textbf{Pix2Pix}    & 0.747908490 & 0.588629693 & 0.988509299 \\
			\hline
			\textbf{CycleGAN}   & 0.376297022 & 0.346849312 & 0.571771837 \\
			\hline
			\textbf{3D-mADUNet} & 1.142354795 & 1.755380266 & 1.875490582 \\
			\hline
			\textbf{PTNet3D}    & 2.796942846 & 0.873885344 & 1.608726699 \\
			\hline
			\textbf{Ea-GAN}     & 0.479778619 & 0.530060148 & 0.594216597 \\
			\hline
			\textbf{MT-Net}     & 1.225086295 & 0.464382954 & 1.542097224 \\
			\hline
			\textbf{ResViT}     & 0.089706534 & 0.060892092 & 0.198871978 \\
			\hline
			\textbf{cWDM}       & 0.261160282 & 0.157456003 & 0.243742557 \\
			\hline
			\textbf{TC-MGAN}    & 0.191847456 & 0.441981262 & 0.760215315 \\
			\hline
			\textbf{3D-MC-SAGAN} & \textbf{0.071416927} & \textbf{0.055989483} & \textbf{0.151343591} \\
			\hline
		\end{tabular}
	\end{table}

	\subsection{Quantitative MRI Synthesis and Downstream Tumour Segmentation}
	
	We evaluate the performance of 3D-MC-SAGAN for multi-contrast MRI synthesis from a single source modality and compare it with alternative methods. At inference, the generator receives a T2w test volume $x_s$ and a target code $c \in {\text{T2f}, \text{T1n}, \text{T1c}}$, producing the corresponding synthesised volume $\hat{y}_c$. Synthesis fidelity is quantified using PSNR, SSIM, and MSE between $\hat{y}_c$ and the paired ground-truth volume $y_c$. Table~\ref{tab:table1} summarises the performance of the proposed method against recent 2D and 3D baselines, with results reported as mean $\pm$ standard deviation over the test set. The results show that 3D-MC-SAGAN achieves the best performance across all target contrasts. Improvements are consistent across PSNR, SSIM, and MSE, indicating reduced voxel-wise error and better structural preservation. These improvements are observed across T2f, T1n, and T1c, demonstrating the robustness and generalisability of the proposed model across different contrast synthesis tasks rather than being limited to a single modality.
	
	The comparative results in Table~\ref{tab:table1} highlight the clear advantages of the proposed 3D-MC-SAGAN over alternative frameworks. 2D slice-wise methods, such as MT-Net, ResViT, and TC-MGAN, can achieve competitive scores, but their lack of through-plane contextual modelling often compromises volumetric coherence, leading to lower PSNR values compared to our method. Patch-based 3D approaches incorporate local volumetric context, yet their limited effective field-of-view can hinder global consistency and introduce boundary artifacts during patch blending, resulting in less reliable structural continuity. Diffusion-based models may perform well under MSE-driven objectives; however, they often produce overly smooth outputs, which can suppress fine contrast details and subtle texture variations. In contrast, 3D-MC-SAGAN effectively preserves global structure and volumetric continuity while maintaining sharper and more accurate anatomical details across all target contrasts.

\begin{figure*}[t]
	\centering
	\includegraphics[width=\textwidth]{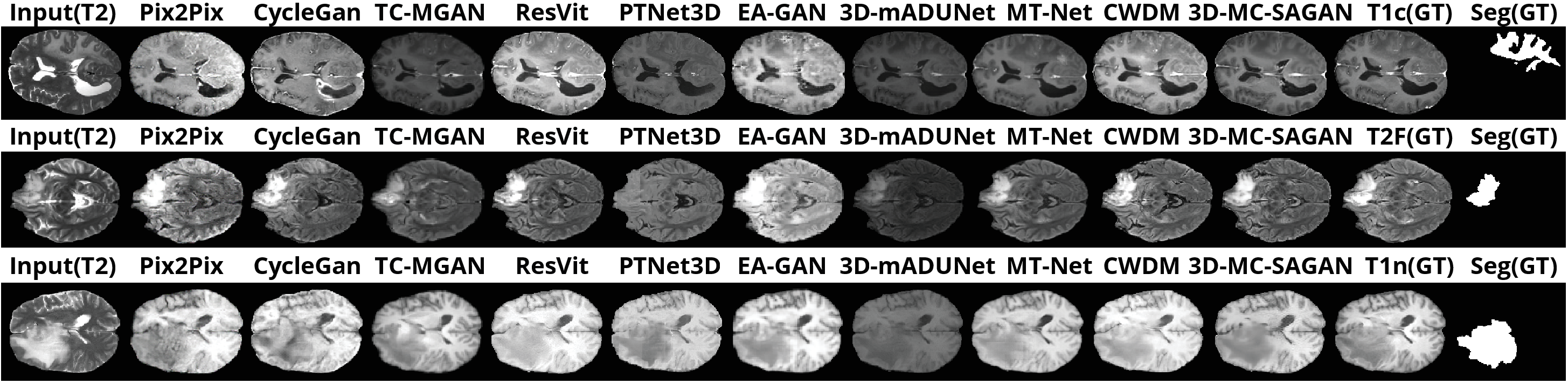}
	\caption{Representative qualitative results for multi-contrast brain MRI synthesis. Rows correspond to target modalities; columns show the source T2w image, comparative results, the 3D-MC-SAGAN result, the real target image, and the binary whole-tumour reference.}
	\label{fig:fig3}
\end{figure*}

	Fig.\ref{fig:fig3} illustrates qualitative results for T2w$\rightarrow$T2f, T1n, and T1c synthesis. These results demonstrate that the proposed 3D-MC-SAGAN consistently produces sharper tissue boundaries, clearer contrast transitions, and better preserves subtle anatomical details, particularly in challenging regions. In contrast, several baseline methods often exhibit visible artifacts and blurred structures. These visual improvements closely reflect the quantitative gains reported in Table\ref{tab:table1}, highlighting the model’s ability to generate anatomically coherent and high-fidelity multi-contrast volumes.
	
	To further evaluate the realism of the generated volumes at the distribution level, we report the MFD for each target modality for both the proposed and alternative methods in Table~\ref{tab:mfd_results}, where lower values indicate closer alignment between generated and real-image feature distributions. The results show that the proposed 3D-MC-SAGAN achieves the lowest MFD across all three modalities, with some baseline methods showing significantly higher values. These results suggest that our model not only preserves contrast-specific appearance statistics more effectively but also ensures improved global consistency beyond voxel-wise similarity, demonstrating its ability to capture the overall structural and textural characteristics of multi-contrast MRI.

	We next evaluate the downstream utility of the generated volumes via tumour segmentation using a fixed segmentation network. For each synthesis method, the three generated modalities (T1n, T1c, T2f) are concatenated with the original T2w input to form a four-channel volume, which is then fed to the segmenter to produce tumour predictions. Dice scores are reported against the reference tumour mask, alongside two real-data baselines: using only T2w, and using all real modalities.
	
	As shown in Table~\ref{tab:table2}, 3D-MC-SAGAN achieves the highest Dice score when using generated modalities (0.8631), slightly higher than all-real-modality baseline (0.8618). The table also shows that several baseline methods exhibit larger drops in Dice, even when their synthesis fidelity appears reasonable. This result suggests that pixel-level similarity alone may not be sufficient to preserve pathology-relevant structure. Methods lacking mechanisms to maintain volumetric consistency often blur tumour boundaries or alter lesion extent. By enforcing global structural and volumetric coherence, 3D-MC-SAGAN more effectively preserves tumour-relevant features while maintaining high synthesis fidelity.

	\begin{table}[t]
		\centering
		\caption{Dice score results. Higher values indicate better segmentation performance.}
		\label{tab:table2}
		\begin{tabular}{|l|c|c|c|}
			\hline
			\textbf{Method} &
			\textbf{\begin{tabular}[c]{@{}c@{}}Dice Score\\(only Real\\ T2w)\end{tabular}} &
			\textbf{\begin{tabular}[c]{@{}c@{}}Dice Score\\(Real Data)\end{tabular}} &
			\textbf{\begin{tabular}[c]{@{}c@{}}Dice Score\\(Generated\\Data)\end{tabular}} \\
			\hline
			\textbf{Pix2Pix} & 0.6975 & 0.8085 & 0.6556 \\
			\hline
			\textbf{CycleGAN} & 0.6975 & 0.8085 & 0.5914 \\
			\hline
			\textbf{3D-mADUNet} & 0.7136 & 0.8618 & 0.8082 \\
			\hline
			\textbf{PTNet3D} & 0.7136 & 0.8618 & 0.5747 \\
			\hline
			\textbf{Ea-GAN} & 0.6998 & 0.8332 & 0.8041 \\
			\hline
			\textbf{MT-Net} & 0.6967 & 0.8068 & 0.7416 \\
			\hline
			\textbf{ResViT} & 0.6975 & 0.8085 & 0.7699 \\
			\hline
			\textbf{cWDM} & 0.7136 & 0.8618 & 0.8418 \\
			\hline
			\textbf{TC-MGAN} & 0.7218 & 0.8047 & 0.7832 \\
			\hline
			\textbf{3D-MC-SAGAN} & 0.7136 & 0.8618 & \textbf{0.8631} \\
			\hline
		\end{tabular}
	\end{table}


	\subsection{Ablation Study}
	We conduct an ablation study to evaluate the contribution of three key components in 3D-MC-SAGAN: MBHA block, the 3D perceptual loss, and the segmentation-consistency loss computed via the frozen segmenter. In each variant, a single component is removed while the remaining architecture, objectives, and training protocol are kept fixed. A reduced baseline that removes all three components is also included. Table~\ref{tab:table3} summarises synthesis fidelity using PSNR, SSIM, and MSE for each target contrast (T2f, T1n, and T1c) and reports Dice to reflect tumour segmentation performance in the synthesised domain.
	
	Across all targets, the full 3D-MC-SAGAN configuration achieves the best balance between synthesis fidelity and segmentation accuracy. Removing MBHA consistently degrades both image quality and Dice, indicating the importance of memory-aware attention for modelling long-range dependencies and maintaining structural coherence. Excluding the perceptual loss results in a smaller but systematic drop in fidelity and Dice, showing that feature-level supervision complements voxel-wise objectives to produce anatomically plausible appearances. The removal of the segmentation-consistency loss leads to the largest reduction in Dice among single ablations and also reduces synthesis metrics, highlighting its key role in preserving tumour morphology during translation. The reduced baseline, which omits all three components, performs worst overall, confirming that these elements are complementary and jointly contribute to robust, pathology-preserving 3D synthesis.

		\renewcommand{\arraystretch}{1}
\begin{table*}[t!]
	\centering
	\setlength{\tabcolsep}{2pt}
	\caption{Ablation study results. PSNR, SSIM, and MSE are reported for T2f, T1n, and T1c. Dice score reflects downstream tumour segmentation performance. Best results are in \textbf{bold}.}
	\label{tab:table3}
	\resizebox{\textwidth}{!}{
		\begin{tabular}{|l|ccc|ccc|ccc|c|}
			\hline
			\multirow{2}{*}{\textbf{Method}} &
			\multicolumn{3}{c|}{\textbf{PSNR}} &
			\multicolumn{3}{c|}{\textbf{SSIM}} &
			\multicolumn{3}{c|}{\textbf{MSE}} &
			\multirow{2}{*}{\textbf{Dice score}} \\ \cline{2-10}
			& \textbf{T2f} & \textbf{T1n} & \textbf{T1c}
			& \textbf{T2f} & \textbf{T1n} & \textbf{T1c}
			& \textbf{T2f} & \textbf{T1n} & \textbf{T1c} 
			& {} \\ \hline
			
			w/o MBHA
			& 26.681$\pm$1.980 & 27.538$\pm$2.087 & 26.031$\pm$2.356
			& 0.940$\pm$0.013 & 0.956$\pm$0.014 & 0.932$\pm$0.014
			& 0.0023$\pm$0.0011 & 0.0020$\pm$0.0011 & 0.0029$\pm$0.0019
			& 0.8472 \\ \hline
			
			w/o Perc.\ Loss
			& 27.069$\pm$2.202 & 27.467$\pm$2.363 & 26.367$\pm$2.267
			& 0.944$\pm$0.014 & 0.960$\pm$0.014 & 0.936$\pm$0.013
			& 0.0022$\pm$0.0012 & 0.0021$\pm$0.0013 & 0.0026$\pm$0.0015
			& 0.8518 \\ \hline
			
			w/o Seg.\ Loss
			& 26.929$\pm$2.344 & 27.449$\pm$2.076 & 25.897$\pm$2.257
			& 0.941$\pm$0.015 & 0.959$\pm$0.013 & 0.932$\pm$0.014
			& 0.0023$\pm$0.0013 & 0.0020$\pm$0.0011 & 0.0029$\pm$0.0016
			& 0.8237 \\ \hline
			
			w/o All above
			& 26.672$\pm$2.436 & 26.951$\pm$2.291 & 25.879$\pm$2.596
			& 0.931$\pm$0.020 & 0.948$\pm$0.017 & 0.924$\pm$0.014
			& 0.0025$\pm$0.0020 & 0.0023$\pm$0.0015 & 0.0031$\pm$0.0020
			& 0.8121 \\ \hline
			
			\textbf{3D-MC-SAGAN}
			& \textbf{27.680$\pm$2.104} & \textbf{28.428$\pm$2.218} & \textbf{26.738$\pm$2.573}
			& \textbf{0.947$\pm$0.014} & \textbf{0.966$\pm$0.013} & \textbf{0.943$\pm$0.015}
			& \textbf{0.0019$\pm$0.0011} & \textbf{0.0016$\pm$0.0009} & \textbf{0.0025$\pm$0.0017}
			& \textbf{0.8631} \\ \hline
			
	\end{tabular}}
\end{table*}

	\section{Conclusion}
	
	In this work, we introduced 3D-MC-SAGAN, a unified framework for synthesising missing multi-contrast brain MRI modalities from a single T2w input while preserving tumour-relevant features. The proposed architecture combines a multi-scale encoder--decoder generator with a novel Memory-Bounded Hybrid Attention (MBHA) block and a Wasserstein-based adversarial formulation to produce high-fidelity reconstructions. A U-Net-style segmentation module provides tumour-consistency guidance through a Dice-driven loss, preserving clinically important lesion features during synthesis. Experimental results on the BraTS 2023 dataset demonstrate that the proposed model generates higher-quality contrasts in tumour regions and supports improved downstream segmentation performance. These findings highlight the potential of 3D-MC-SAGAN to reduce the need for exhaustive MRI acquisition while maintaining diagnostically meaningful information.

\end{document}